\documentstyle{article}
\title{Green functions and Euclidean fields near the bifurcate Killing horizon }
\author{ Z. Haba\\Institute of Theoretical Physics, University of Wroclaw,
\\50-204 Wroclaw, Plac Maxa Borna 9, Poland\\e-mail:zhab@ift.uni.wroc.pl}
\date{PACS numbers 04.62+v,04.70.Dy }
\begin{document}
\maketitle
\begin{abstract}
We approximate  a Euclidean version of a $D+1$ dimensional
manifold with a bifurcate Killing horizon by a product of the two
dimensional Rindler space ${\cal R}_{2}$  and a $D-1$ dimensional
Riemannian manifold ${\cal M}$. We obtain approximate formulas for
the Green functions. We study the behaviour of Green functions
near the horizon and their dimensional reduction.
 We show that if ${\cal M}$ is compact
 then  the massless minimally coupled
quantum  field contains a zero mode which is a conformal invariant
free field  on $R^{2}$. Then, the Green function near the horizon
can be approximated by
 the Green function of the two-dimensional quantum field theory.
 The correction term is exponentially small  away from the
 horizon. If the volume of a geodesic ball is growing to infinity with its
 radius then the Green function cannot be approximated by a
 two-dimensional one.
 \end{abstract}
\section{Introduction}
We are interested in a study of the quantum version of the phenomena associated
with a motion of a particle around the black hole.
As the quantum mechanics in an external field encounters
the problem of particle creation  leading to many particle systems
we feel that a proper approach to the background gravitational field
goes through the field quantization.
Quantum field theory can be defined by means of Green functions.
 In the Minkowski space the locality and Poincare invariance
 determine the Green functions and allow a construction of free quantum fields.
  In the curved space  the Green function is not unique. The
non-uniqueness can be interpreted as a non-uniqueness of the
physical vacuum \cite{fulling}\cite{davis}. There is less
ambiguity in the definition of the Green function on the
Riemannian manifolds (instead of the physical pseudo-Riemannian
ones).  The Euclidean approach appeared successful when applied to
the construction of quantum fields on the Minkowski spacetime
\cite{jaffe0}. We hope that such an approach will be fruitful in
application to a curved background as well. In contradistinction
to the Minkowski spacetime an analytic continuation of Euclidean
fields to quantum fields from the Riemannian metric to the
pseudoRiemannian one is possible only if the manifold has an
additional reflection symmetry \cite{dimock}\cite{jaffe}. The
reflection symmetry may have a physical meaning concerning
tunneling phenomena which could justify the requirement of an
additional symmetry of the gravitational background
\cite{gibbons}. The Euclidean approach to quantum fields on a
curved background has been discussed earlier in \cite{wald0}
\cite{angelis}\cite{haba0} and developed in
\cite{dimock}\cite{jaffe}.

 The event horizon has a crucial
relevance for locality of quantum field theory because some
information is lost behind the horizon. It is also the defining
property of the black hole. The event horizon is a global property
of the pseudoRiemannian manifold. Hence, it is hard to see how it
could be defined after an analytic continuation to the Riemannian
manifold. There is however a proper substitute:the bifurcate
Killing horizon \cite{wald2}. As proved in \cite{racz} a manifold
with the Killing horizon ( a static black hole is an example of
the Killing horizon) can be extended to the manifold with a
bifurcate Killing horizon. Moreover, there always exists an
extension with the wedge reflection symmetry \cite{kay} which
seems crucial for  quantum field theory and for an analytic
continuation between pseudoRiemannian and Riemannian manifolds.
 The bifurcate Killing horizon is a local
property which can be treated in local coordinates
\cite{wald2}. In local Kruskal-Szekeres type of coordinates close to the
bifurcate Killing horizon the metric tensor $g_{\mu\nu}$
 tends to zero at the horizon. This property
is preserved after a continuation to the Riemannian metric. We can
treat approximately the Riemannian manifold ${\cal N}$  with the
bifurcate Killing horizon  as ${\cal N}={\cal R}_{2}\times {\cal
M}_{D-1}$, where ${\cal R}_{2}$ is the two dimensional Rindler
space and ${\cal M}_{D-1}$ enters the definition of the bifurcate
Killing horizon as an intersection of past and future horizons.
There are well-known examples of the approximation in the form of
a product: the Schwarzschild solution can be approximated near the
horizon by a product of the Rindler space and  the two-dimensional
sphere . However, we do not restrict ourselves to metrics which
are solutions of Einstein equations .

 We consider an
equation for the Green functions in an approximate metric near the
bifurcate Killing horizon . We expand the solution into
eigenfunctions of the Laplace-Beltrami operator on ${\cal M}$. If
${\cal M}$ is compact without a boundary then the Laplace
-Beltrami operator has a discrete spectrum starting from $0$ (the
zero mode). We show that the higher modes are damped by a
tunneling mechanism. As a consequence the position of the point
defining the Green function on the manifold ${\cal M}$ becomes
irrelevant. The Green function near the bifurcate Killing horizon
can be well approximated by the Green function of the
two-dimensional conformal free field. The splitting of the Green
function near the horizon into a product of the two dimensional
function and a function on ${\cal M}$ has been predicted by
Padmanabhan \cite{padma}. However, we obtain its exact form.
Moreover, we show that if ${\cal M}$ is not compact, but the
volume of a geodesic ball of radius $r$ grows like a power of $r$,
then the reduction to a two dimensional Green function does not
take place.

The mechanism of the dimensional reduction presented in this paper
may work for other models. In particular, for the ones with a
warped form of the metric \cite{warp} encountered in the brane
models \cite{warpbrane}. Such models could explain why the Universe
is confined to a submanifold and the probability of tunneling out of it
is small.

\section{An approximation at the horizon}We consider a $D+1$ dimensional Riemannian
manifold ${\cal N}$ with a metric $g_{AB}$ characterized by a
bifurcate Killing horizon. This notion assumes a symmetry
generated by the Killing vector $\xi^{A}$. Then, it is assumed
that the Killing vector is orthogonal to a (past oriented) $D$
dimensional hypersurface ${\cal H}_{A}$ and a (future oriented)
hypersurface ${\cal H}_{B}$ \cite{wald2}. The Killing vector
$\xi^{A}$ is vanishing (i.e.,$\xi^{A}(x)\xi_{A}(x)=0$) on an
intersection of ${\cal H}_{A}$ and ${\cal H}_{B}$ defining a $D-1$
dimensional surface ${\cal M}$ (which can be described as the
level surface $f(x)=const$) .  The bifurcate Killing horizon
implies that the space-time has locally a structure of the one
seen from an accelerated frame, i.e., the structure of the Rindler
space.  Padmanabhan \cite{padma} describes such a bifurcate
Killing horizon as a transformation from a local Lorentz frame to
the local accelarated (Rindler) frame. In \cite{racz} it is proved
that the space-time with a Killing horizon can be extended to a
space-time with the bifurcate Killing horizon. There, it is also
shown that the extension can be chosen in such a way that the
"wedge reflection symmetry" \cite{kay} is satisfied.In the local
Rindler coordinates  the reflection symmetry is $(x_{0},y,{\bf
x})\rightarrow (x_{0},-y,{\bf x})$. The symmetry means that the
metric splits into a block form
\begin{displaymath}
ds^{2}=\sum_{a,b=0,1}g_{ab}dx^{a}dx^{b}+\sum_{jk>1}g_{jk}dx^{j}dx^{k}
\end{displaymath}
 The bifurcate Killing horizon
 distinguishes
a two-dimensional subspace of the tangent space.  At the bifurcate
Killing horizon the two-dimensional metric tensor $g_{ab}$ is
degenerate. In the adapted coordinates such that
$\xi^{A}=\partial_{0}$ we have $g_{10}=0$ and the metric does not
depend on $x_{0}$. Then, $\det[g_{ab}]\rightarrow 0$ at the
horizon means that $g_{00}(y=0,{\bf x})=0$ or $g_{11}(y=0,{\bf
x})=0$ (if both of them were zero then the curvature tensor would
be singular). We assume $g_{00}(y=0,{\bf x})=0$. As $g_{00}$ is
non-negative its Taylor expansion must start with $y^{2}$. Hence,
if we neglect the dependence of the two-dimensional metric
$g_{ab}$ on  ${\bf x}$ then we can write it in the form
\begin{equation} \begin{array}{l}ds_{g}^{2}\equiv
g_{AB}dx^{A}dx^{B}=-y^{2}(dx^{0})^{2}+dy^{2}+\sum_{j,k\geq
2}g_{jk}(y,{\bf x})dx^{j}dx^{k}\cr \equiv y^{2}
\Big(-(dx^{0})^{2}+y^{-2}(dy^{2}+ds_{D-1}^{2})\Big)
\end{array}\end{equation}
If we neglect the dependence of $g_{jk}$ on $y$  near the horizon
then the metric $ds_{D-1}^{2}$ ( denoted $ds_{M}^{2}$ ) can be
considered as a metric
 on the $D-1$ dimensional  surface ${\cal M}$  being the common part of
 ${\cal H}_{A}$ and ${\cal H}_{B}$.  Hence, in eq.(1)
 ${\cal N}={\cal R}_{2}\times {\cal M}$ where ${\cal R}_{2}$
is the two-dimensional Rindler space. As an example of the
approximation of the geometry of ${\cal N}$
 we could consider the four dimensional Schwarzschild black hole when  ${\cal N}\simeq
 {\cal R}_{2}\times
 S^{2}$ ( quantum theory with such an approximation is discussed in \cite{bykowski})  .

We shall work with  Euclidean  version of the metric (1)
\begin{equation} \begin{array}{l}ds^{2}
=y^{2}(dx^{0})^{2}+dy^{2}+\sum_{j,k\geq 2}g_{jk}(y,{\bf
x})dx^{j}dx^{k}\cr \equiv y^{2}
\Big((dx^{0})^{2}+y^{-2}(dy^{2}+ds_{D-1}^{2})\Big)
\end{array}\end{equation}
It is assumed that this is a Riemannian metric on the Riemannian
section of a certain complexified manifold ${\cal N}$
\cite{chrusciel}.

Let
\begin{equation}
\triangle_{N}=\frac{1}{\sqrt{g}}\partial_{A}g^{AB}\sqrt{g}\partial_{B}
\end{equation}
be the Laplace-Beltrami operator on ${\cal N}$.

We are interested in the calculation of the Green functions
\begin{equation}
(-\triangle_{N} +m_{N}^{2}){\cal
G}_{N}^{m}=\frac{1}{\sqrt{g}}\delta
\end{equation}
 A solution of eq.(4) can be expressed by the fundamental solution
 of the diffusion equation
\begin{displaymath}
\partial_{\tau}P=\frac{1}{2}\triangle_{N}P
\end{displaymath}
with the initial condition
$P_{0}(x,x^{\prime})=g^{-\frac{1}{2}}\delta(x-x^{\prime})$. Then
\begin{equation}
{\cal
G}_{N}^{m}=\frac{1}{2}\int_{0}^{\infty}d\tau\exp(-\frac{1}{2}m_{N}^{2}\tau)
P_{\tau}
\end{equation}
In order to prove Eq.(5) we multiply the diffusion equation by
$\exp(-\frac{1}{2}m_{N}^{2}\tau)$ and integrate both sides over
$\tau$ applying the initial condition for $P_{\tau}$.

In the approximation (2) we have (if $g_{jk}$ is independent of $y$)
\begin{displaymath}
 \triangle_{N}=y^{-2}\partial_{0}^{2}+y^{-1}\partial_{y}y\partial_{y}+\triangle_{M}
 \end{displaymath}
where $\triangle_{M}$ is the Laplace-Beltrami operator for the
metric
\begin{displaymath}
ds_{M}^{2}=\sum_{jk}g_{jk}(0,{\bf x})dx^{j}dx^{k}
\end{displaymath}
Then, eq.(4) reads
\begin{equation}
-(\partial_{0}^{2}+y\partial_{y}y\partial_{y}+y^{2}\triangle_{M}-y^{2}m_{N}^{2}){\cal
G}_{N}^{m}=y\frac{1}{\sqrt{g}_{M}}\delta
\end{equation}
 After  an exponential change of coordinates
\begin{equation}
y=b\exp u
\end{equation}
eq.(6) takes the form
\begin{equation}
\Big(-\partial_{0}^{2}-\partial_{u}^{2}-b^{2}\exp(2u)(\triangle_{M}-m_{N}^{2})\Big){\cal
G}^{m}_{N}=g_{M}^{-\frac{1}{2}}\delta(x_{0}-x_{0}^{\prime})\delta(u-u^{\prime})\delta({\bf
x}-{\bf x}^{\prime})
\end{equation}

Let us denote
\begin{equation}
\int d{\bf x}\sqrt{g_{M}}{\cal G}_{N}^{0}\equiv
G(x_{0},x_{1};x_{0}^{\prime},x_{1}^{\prime})
\end{equation}Then, the integrated Green functions (8)-(9)  with $m_{N}=0$
are solutions of the equation for the two-dimensional Green
function
\begin{equation}
(-\partial_{0}^{2}-\partial_{1}^{2})G=\delta(x_{0}-x_{0}^{\prime})\delta(x_{1}-x_{1}^{\prime})
\end{equation}

\section{Generalized hyperbolic manifold ${\cal D}$}

 If ${\cal M}=R^{n}$ ($n=D-1$ ) then we
have in eq.(2) the Rindler metric \cite{rindler}
 \begin{equation}
 ds^{2}=y^{2}(dx^{0})^{2}+dy^{2}+(dx^{2})^{2}+....+(dx^{D})^{2}
 \end{equation}
 where $y>0$ and ${\bf x}\in R^{D-1}$.
In the Rindler $D+1$ dimensional space the formula (3) for the
Laplace-Beltrami operator  reads
\begin{equation}
 \triangle_{R}=y^{-2}\partial_{0}^{2}+y^{-1}\partial_{y}y\partial_{y}+\partial_{2}^{2}+....+
 \partial_{D}^{2}
 \end{equation}
In \cite{haba} we have  expressed the massless Green function for
Rindler space by the massive one for the hyperbolic space. The
conformal relation between metrics is responsible for a relation
between Green functions. In the same way let us consider a
conformal transformation of the metric on ${\cal N}$ to the
optical metric
 \cite{op} $\tilde{g}_{AB}\equiv
g_{AB}g_{00}^{-1}$ when
\begin{equation}
ds_{op}^{2}=
\tilde{g}_{AB}dx^{A}dx^{B}=(dx^{0})^{2}+y^{-2}(dy^{2}+ds_{M}^{2})
\end{equation}
The equation for the Green function in the metric (13) is
\begin{equation}
-\Big(\partial_{0}^{2}+y^{2}\partial_{y}^{2}-(D-2)y\partial_{y}+y^{2}\triangle_{M}
-m^{2}\Big){\cal G}_{op}=y^{D}g_{M}^{-\frac{1}{2}}\delta
\end{equation}
If we introduce ($n= D-1$)
\begin{equation}{\cal G}_{op}(X,X^{\prime})=y^{\frac{n}{2}}y^{\prime\frac{n}{2}}\hat{{\cal G}}_{op}(X,X^{\prime})
\end{equation}
then \begin{equation}
-\Big(\partial_{0}^{2}+y\partial_{y}y\partial_{y}-\frac{n^{2}}{4}+y^{2}\triangle_{M}
-m^{2}\Big)\hat{{\cal G}}_{op}=y g_{M}^{-\frac{1}{2}}\delta
\end{equation}
Comparing eqs.(16) and eq.(6) we can see that if $m_{N}=0$ and
$m^{2}=-\frac{n^{2}}{4}$ then ${\cal G}_{N}$ coincides with ${\cal
G}_{op}$.

The form of the optical metric (13) suggests that we should study a
$D$ dimensional manifold ${\cal D}=R\times {\cal M}$ with the metric
\begin{equation}
ds_{D}^{2}= y^{-2}(dy^{2}+ds_{M}^{2})
\end{equation}
The equation (4) for the Green function on ${\cal D}$ reads

\begin{equation}
-\Big(y^{2}\partial_{y}^{2}-(D-2)y\partial_{y}+y^{2}\triangle_{M}
-m^{2}\Big){\cal G}_{D}^{m}=y^{D}g_{M}^{-\frac{1}{2}}\delta
\end{equation}
or after the transformation (15)
\begin{equation}
-\Big(y\partial_{y}y\partial_{y}-\frac{(D-1)^{2}}{4}+y^{2}\triangle_{M}
-m^{2}\Big)\hat{{\cal G}}_{D}=y g_{M}^{-\frac{1}{2}}\delta
\end{equation}

We consider the Fourier transform of the Green function (6)
\begin{equation}
{\cal G}_{N}^{m}(x_{0},u,{\bf x};x_{0}^{\prime},u^{\prime},{\bf
x}^{\prime})=\int
dp_{0}\exp(ip_{0}(x_{0}-x_{0}^{\prime}))\tilde{{\cal
G}}_{N}^{m}(p_{0},u,{\bf x};u^{\prime},{\bf x}^{\prime})
\end{equation}
It follows from eqs.(6) and (19) that at $m_{N}=0$
\begin{equation}
\tilde{{\cal G}}_{N}^{0}(p_{0},u,u^{\prime};{\bf x},{\bf
x}^{\prime})=\hat{{\cal G}}_{D}^{m}(u,u^{\prime}; {\bf x},{\bf
x}^{\prime})
\end{equation}
if
\begin{equation}
p_{0}^{2}=m^{2}+\frac{n^{2}}{4}
\end{equation}
We define the operator
\begin{equation}
{\cal
B}=y\partial_{y}y\partial_{y}-\frac{(D-1)^{2}}{4}+y^{2}\triangle_{M}
\end{equation}
and its heat kernel $\hat{P}$ (this heat kernel is discussed also
in \cite{bykowski})
\begin{equation}
\partial_{\tau}\hat{P}_{\tau}=\frac{1}{2}{\cal B}\hat{ P}_{\tau}
\end{equation}
with the initial condition
$\hat{P}_{0}(X,X^{\prime})=y\frac{1}{\sqrt{g_{M}}}\delta(X-X^{\prime})$
(here $X=(y,{\bf x})$).

If we have the fundamental solution (24) then integrating over
$\tau$ we can solve the equation for the Green function
\begin{equation}
(-{\cal B}+ m^{2}){\cal G}_{D}^{m}=y\frac{1}{\sqrt{g}}\delta
\end{equation}
Applying eqs.(21),(24) and (25) we obtain \begin{equation}
\tilde{{\cal G}}_{N}^{0}(p_{0},u,u^{\prime}; {\bf x},{\bf
x}^{\prime})
=\int_{0}^{\infty}d\tau\exp(-\frac{\tau}{2}p_{0}^{2}+\frac{\tau}{8}n^{2})\hat{P}_{\tau}
(y,{\bf x};y^{\prime},{\bf x}^{\prime})
\end{equation}
or
\begin{equation}
{\cal G}_{N}^{0}(x_{0},u,{\bf x};x_{0}^{\prime},u^{\prime},{\bf
x}^{\prime})=\int_{0}^{\infty}d\tau(2\pi\tau)^{-\frac{1}{2}}
\exp(-\frac{1}{2\tau}(x_{0}-x_{0}^{\prime})^{2}+\frac{\tau}{8}n^{2})\hat{P}_{\tau}
(y,{\bf x};y^{\prime},{\bf x}^{\prime})
\end{equation}

\section{The heat kernel on $ {\cal D} $}

The Laplace-Beltrami operator $\triangle_{D}$ on ${\cal D}$ reads
\begin{equation}
\triangle_{D}=y^{2}\partial_{y}^{2}-(D-2)y\partial_{y}+y^{2}\triangle_{M}
 \end{equation}
We consider the heat equation
\begin{equation}
\partial_{\tau}P_{\tau}=\frac{1}{2}\triangle_{D}P_{\tau}
\end{equation}
with the initial condition $y^{-D}g_{M}^{-\frac{1}{2}}\delta$.

If ${\cal M}$ is a compact manifold without a boundary then the
spectrum of the  Laplace-Beltrami operator $\triangle_{M}$ is
discrete \cite{chavel}
\begin{equation}
-\triangle_{M}u_{k}=\epsilon_{k}u_{k}
\end{equation}
 The eigenfunctions $u_{k}$ are normalized
\begin{displaymath}
\int d{\bf x}\sqrt{g}_{M}\overline{u}_{n}u_{k}=\delta_{nk}
\end{displaymath}
and satisfy the completeness relation
\begin{equation}
\sum_{k}\overline{u}_{k}({\bf x})u_{k}({\bf
x}^{\prime})=g_{M}^{-\frac{1}{2}}\delta({\bf x}-{\bf x}^{\prime})
\end{equation}
Let us note that on a compact manifold a constant is an
eigenfunction (30) with the lowest eigenvalue $0$ as
\begin{displaymath}
\triangle_{M}1=0
\end{displaymath}
This zero mode is crucial for the behaviour of the Green function
near the bifurcate Killing horizon.

We can expand the fundamental solution of the heat equation (29) in
 the complete set of eigenfunctions (30)
\begin{equation}
P_{\tau}(y,{\bf x};y^{\prime},{\bf x}^{\prime})=\sum_{k}q_{\tau}^{k}(y,y^{\prime})\overline{u}_{k}({\bf x})u_{k}({\bf
x}^{\prime})
\end{equation}
Inserting in eq.(29) we obtain an ordinary differential equation for $q_{\tau}^{k}$.

 In the Appendix we
calculate $P_{\tau}$ assuming that $P_{\tau}$ depends only on the
geodesic distance $\sigma$ on ${\cal D}$. First, we show that if
$\sigma_{M}$ is the geodesic distance on ${\cal M}$ then $\sigma$ is
determined by the formula (this is a generalization of the
well-known formula for the hyperbolic space)
\begin{equation}
\cosh\sigma=1+(2yy^{\prime})^{-1}(\sigma_{M}^{2}+(y-y^{\prime})^{2})
\end{equation}
In order to prove eq.(33) we  show by means of a direct
calculation that $\sigma$ satisfies the Hamilton-Jacobi equation
\cite{dewitt}
\begin{equation}
y^{2}(\partial_{y}\sigma\partial_{y}\sigma+g^{jk}\partial_{j}
\sigma\partial_{k}\sigma) =1
\end{equation}
if $\sigma_{M}$ satisfies the equation
 \begin{equation}
g^{jk}\partial_{j}\sigma_{M}
\partial_{k}\sigma_{M}
=1
\end{equation}
 We return to the heat equation (24) which results by the transformation
 (15) from the heat equation (29) . We integrate both sides of eq.(24) with respect to the
measure $\sqrt{g_{M}}d{\bf x}$ of the manifold ${\cal M}$. Denote
\begin{equation}
q_{\tau}(y,y^{\prime})=\int\sqrt{g_{M}}d{\bf
x}\hat{P}_{\tau}(y,{\bf x};y^{\prime},{\bf x}^{\prime})
\end{equation}
It follows that $q_{\tau}$ satisfies the equation
\begin{equation}
\partial_{\tau}q_{\tau}=\frac{1}{2}\Big(\partial_{u}^{2}-\frac{(D-1)^{2}}{4}\Big)q_{\tau}
\end{equation}
with the initial condition $q_{0}=\delta(u-u^{\prime})$ (this is the same equation as for
$q_{\tau}^{0}$ in eq.(32) after an exponential change of coordinates (7)).

The solution of eq.(37) is
\begin{equation}
q_{\tau}(u,u^{\prime})=(2\pi\tau)^{-\frac{1}{2}}\exp\Big(-\frac{1}{2\tau}(u-u^{\prime})^{2}
-\frac{\tau}{8}(D-1)^{2}\Big)
\end{equation}When the lhs of eq.(36) is known (from eq.(38))
then eq.(36) can be considered as an integral equation for
$\hat{P}_{\tau}$.

We can check using eqs.(27) and (38) that eq.(10) is satisfied as
\begin{equation}\begin{array}{l}
\int d{\bf x}\sqrt{g_{M}}{\cal G}_{N}^{0}(y,x_{0},{\bf
x};y^{\prime},x_{0}^{\prime},{\bf x}^{\prime})=\int_{0}^{\infty}
d\tau (2\pi\tau)^{-1}\exp(-\frac{1}{2\tau}(u-u^{\prime})^{2}
-\frac{1}{2\tau}(x_{0}-x_{0}^{\prime})^{2})\cr =-\frac{1}{4\pi}
\ln\Big((u-u^{\prime})^{2} +(x_{0}-x_{0}^{\prime})^{2}\Big)
\end{array}\end{equation}

 Let us define the geodesic ball at the point ${\bf x}_{0}$
\begin{displaymath}
B_{r}({\bf x}_{0})=({\bf y}\in {\cal M}:\sigma({\bf x}_{0},{\bf
y})\leq r )
\end{displaymath}We denote
\begin{equation}
\Omega(r)=Vol( B_{r}({\bf 0}))\equiv\int_{0}^{r}d\rho\omega(\rho)
\end{equation} defining the density $\omega(r)$.

The volume element is\begin{equation} d{\bf x}\sqrt{g_{M}}({\bf
x})=dr\omega(r)dS
\end{equation}
where $dS$ denotes an integral over the surface of the ball (with
a  normalization $\int dS=1$).
 Then, under the assumption that $P_{\tau}$ depends
  only on the geodesic distance, eq.(36) (rewritten in terms of $P$ of eq.(29)) reads
 \begin{equation}
 \begin{array}{l}
 \int_{0}^{\infty}dr\omega(r)P_{\tau}\Big((2yy^{\prime})^{-1}(r^{2}+(y-y^{\prime})^{2})\Big)
 \cr
 =(yy^{\prime})^{\frac{n}{2}}(2\pi
 \tau)^{-\frac{1}{2}}\exp\Big(-(2\tau)^{-1}(\ln y - \ln
 y^{\prime})^{2}-\frac{n^{2}\tau}{8}\Big)
 \end{array}
 \end{equation}
 As an example: heat kernels on homogeneous spaces of rank 1
 depend only on the geodesic distance \cite{campo}\cite{gaveau}.
 In general, we must treat our assumption as an approximation.

  Let us denote
 \begin{equation}
 \rho=\frac{r^{2}}{2yy^{\prime}}
 \end{equation}
 Then, eq.(42) can be rewritten as
 \begin{equation}
 \int_{0}^{\infty}P_{\tau}(\rho+v)f(2yy^{\prime}\rho)
 d\rho=T_{n}(\tau,v)
 \end{equation}
 where
 \begin{equation}
 f(u)=\omega(\sqrt{u})u^{-\frac{1}{2}}\end{equation}
and
 \begin{displaymath}
 v=(2yy^{\prime})^{-1}(y-y^{\prime})^{2}
 \end{displaymath}
 can be considered as $\cosh \sigma -1$ at ${\bf x}={\bf
 x}^{\prime}$. The rhs of eq.(44)
 \begin{equation}
 T_{n}(\tau,v)=\frac{1}{2}(yy^{\prime})^{\frac{n-2}{2}}(2\pi
 \tau)^{-\frac{1}{2}}\exp\Big(-\frac{1}{2\tau}w^{2}-\frac{n^{2}\tau}{8}\Big)
 \end{equation}
 is  a function of
 \begin{displaymath}
 w=\ln\frac{y^{\prime}}{y}
 \end{displaymath}
 or a function of
 \begin{equation}
v=\cosh w -1
\end{equation}
Then,  $w$ in eq.(47) is replaced by  the geodesic distance, so
that in eq.(44)
 $v=\cosh\sigma -1$ .

When the function $\omega $ is known then we obtain an integral
equation for $\hat{P}$. On $R^{n}$
\begin{equation}
\omega(r)=A(n-1)r^{n-1}
\end{equation}
where $A(n-1)$ is the area of an $n-1$ dimensional unit sphere.

We can see from eq.(44)-(45) that $P_{\tau}$ is determined by the
volume measure $\Omega$. We solve eq.(44) for $P_{\tau}$ in the
Appendix under the assumption that the volume $\Omega(r)$ grows
like a power of $r$ . For compact manifolds we study the behaviour
of $P_{\tau}$ in the next section.

\section{Euclidean free fields near  the horizon}
We investigate in this section the Green function (8) in $D+1$
dimensions under the assumption that ${\cal M}$ is $D-1$
dimensional compact manifold without a boundary with a complete
set of eigenfunctions (30)-(31). We introduce the complete basis
of eigenfunctions in the space $L^{2}(dx_{0}dx_{1})$ of the
remaining two  coordinates
\begin{equation}
(-\partial_{0}^{2}-\partial_{1}^{2}+b^{2}\epsilon_{k}\exp(2x_{1})\Big)\phi_{k}^{E}(x_{0},x_{1})=E
\phi_{k}^{E}(x_{0},x_{1})
\end{equation}
In eq.(49) $E$ denotes the set of all the parameters the solution
$\phi^{E}$ depends on. The solutions $\phi$ satisfy the
completeness relation
\begin{equation}
\int
d\nu(E)\overline{\phi}_{k}^{E}(x_{0},x_{1})\phi_{k}^{E}(x_{0}^{\prime},x_{1}^{\prime})
=\delta(x_{0}-x_{0}^{\prime })\delta(x_{1}-x_{1}^{\prime })
\end{equation}
with a certain measure $\nu$ and the orthogonality relation
\begin{equation}\int
dx_{0}dx_{1}\overline{\phi}_{k}^{E}(x_{0},x_{1})\phi_{k}^{E^{\prime}}(x_{0},x_{1})=
\delta(E-E^{\prime})\end{equation} where again the $\delta$
function concerns all parameters characterizing the solution. Then,
we expand the Green function into the Kaluza-Klein modes $u_{k}$ as in eq.(32)
\begin{equation}\begin{array}{l}
{\cal G}_{N}(x_{0},x_{1},{\bf
x};x_{0}^{\prime},x_{1}^{\prime},{\bf x}^{\prime})\cr =\sum_{k}
g_{k}(x_{0},x_{1};x_{0}^{\prime},x_{1}^{\prime})
\overline{u}_{k}({\bf x})u_{k}({\bf
x}^{\prime})\end{array}\end{equation}where
\begin{equation}
g_{k}(x_{0},x_{1};x_{0}^{\prime},x_{1}^{\prime})=\int
d\nu(E)E^{-1}\overline{\phi}_{k}^{E}(x_{0},x_{1})\phi_{k}^{E}(x_{0}^{\prime},x_{1}^{\prime})
\end{equation}
$g_{k}$ is the kernel of the inverse of the operator
\begin{displaymath}
H_{k}=-\partial_{0}^{2}-\partial_{1}^{2}+b^{2}\epsilon_{k}\exp(2x_{1})
\end{displaymath}
i.e.,
\begin{displaymath}
H_{k}g_{k}(x_{0},x_{1};x_{0}^{\prime},x_{1}^{\prime})=\delta(x_{0}-x_{0}^{\prime})\delta(x_{1}-x_{1}^{\prime})
\end{displaymath}
The integral over ${\bf x}$  eliminates all $u_{k}$ from the sum
(52) except of $k=0$. Hence,
\begin{equation}\begin{array}{l}
 G(x_{0},x_{1};x_{0}^{\prime},x_{1}^{\prime})=\int d\nu(E) E^{-1}
\overline{\phi}_{0}^{E}(x_{0}^{\prime},x_{1}^{\prime})\phi_{0}^{E}(x_{0},x_{1})\cr=
-\frac{1}{4\pi}\ln \Big((x_{0}-x_{0}^{\prime
})^{2}+(x_{1}-x_{1}^{\prime })^{2}\Big)\end{array}\end{equation}
In the  limit
$x_{1}\rightarrow +\infty$ the eigenfunctions $\phi^{E}$ decay
exponentially (except of the zero mode) when entering the
potential barrier. Hence, we may expect that  the Green functions also
 decay exponentially from the zero mode contribution ${\cal G}_{0}$ away from
the horizon .

We study this phenomenon in more detail now. First, we write the
solution of eq.(49) in the form
\begin{equation}
\phi_{k}^{E}=\exp(ip_{0}x_{0})\phi_{k}^{p_{1}}(x_{1})
\end{equation}where

\begin{equation}
(-\partial_{1}^{2}+b^{2}\epsilon_{k}\exp(2x_{1}))\phi_{k}^{p_{1}}=p_{1}^{2}\phi_{k}^{p_{1}}
\end{equation}Now, $E=p_{0}^{2}+p_{1}^{2}$ and $d\nu=dp_{0}dp_{1}$ in
eqs.(49)-(51). When $\epsilon_{k}=0$ then the solution of eq.(56)
is the plane wave
\begin{displaymath}
\phi_{0}^{p_{1}}=(2\pi)^{-\frac{1}{2}}\exp(ip_{1}x_{1})
\end{displaymath}
Hence, eq.(54) follows by an explicit calculation.

 The normalized
solution of eq.(56) which behaves like a plane wave with momentum
$p_{1}$ for $x_{1}\rightarrow -\infty$ and decays exponentially
for $x_{1}\rightarrow +\infty$ reads
\begin{equation}
\phi_{k}^{p_{1}}=N_{p_{1}}K_{ip_{1}}(b\sqrt{\epsilon_{k}}\exp(x_{1}))
\end{equation}
where $K_{\nu}$ is the modified Bessel function of the third kind
of order $\nu$ \cite{grad}.

 This solution is inserted into the formula (52)
for the Green function with the normalization (51)
\begin{equation}
\int_{-\infty}^{\infty}dx_{1}\overline{\phi}_{k}^{p_{1}}(x_{1})\phi_{k}^{p_{1}^{\prime}}(x_{1})
=\delta(p_{1}-p_{1}^{\prime})
\end{equation}
Hence (see \cite{sommerfield}),  \begin{equation} N_{p_{1}}^{2}
=p_{1}\sinh(\pi p_{1})\frac{2}{\pi^{2}}
\end{equation} Then,  performing the integral over $p_{0}$ in
eq.(53)
\begin{displaymath}
\int dp_{0}\exp(ip_{0}(x_{0}-x_{0}^{\prime}))(p_{0}^{2}+p_{1}^{2})^{-1}=
\pi \vert p_{1}\vert^{-1}\exp(-\vert p_{1}\vert \vert x_{0}-x_{0}^{\prime}\vert)
\end{displaymath}
we obtain ($G$ is defined in eq.(54))
\begin{equation}\begin{array}{l}
{\cal G}_{N}(x_{0},x_{1},{\bf
x};x_{0}^{\prime},x_{1}^{\prime},{\bf x}^{\prime})-
G(x_{0},x_{1};x_{0}^{\prime},x_{1}^{\prime})
=\frac{4}{\pi^{2}}\int_{0}^{\infty}dp_{1}\sinh(\pi
p_{1})\exp(-p_{1}\vert x_{0}-x_{0}^{\prime}\vert) \cr\sum_{k\neq
0}K_{ip_{1}}(b\sqrt{\epsilon_{k}}\exp(x_{1}))
K_{ip_{1}}(b\sqrt{\epsilon_{k}}\exp(x_{1}^{\prime}))
\overline{u}_{k}({\bf x})u_{k}({\bf
x}^{\prime})\end{array}\end{equation}  An estimate of the sum over
$k$ is difficult in general. The sum itself should be understood
in the sense of a convergence of bilinear forms ( or equivalently
in the sense of a convergence of the partial sum as a
distribution). There are some subtleties in this convergence. For
example, if we let $b\rightarrow 0$ then the solution of eq.(8)
($u=x_{1} , m_{N}=0$) is
\begin{displaymath}
{\cal G}_{N}(x_{0},x_{1},{\bf
x};x_{0}^{\prime},x_{1}^{\prime},{\bf
x}^{\prime})=G(x_{0},x_{1};x_{0}^{\prime},x_{1}^{\prime})
g_{M}^{-\frac{1}{2}}\delta({\bf x}-{\bf x}^{\prime})
\end{displaymath}
The limit $x_{1}\rightarrow -\infty$ could possibly be realized as $b\rightarrow 0$.
In the latter limit the rhs of eq.(60) should tend to
\begin{displaymath}
G(x_{0},x_{1};x_{0}^{\prime},x_{1}^{\prime})g_{M}^{-\frac{1}{2}}\sum_{k\neq
0}\overline{u}_{k}({\bf x})u_{k}({\bf x}^{\prime})
\end{displaymath}
It is difficult to prove such limits in general.

We shall restrict ourselves to particular examples and heuristic
arguments for some estimates of the rhs of eq.(60) which however
cannot be uniform in $x_{1}\rightarrow -\infty$. Let us consider
the simplest example ${\cal M}=S^{1}$. Then,
$u_{k}(x_{2})=(2\pi)^{-\frac{1}{2}}\exp(ikx_{2})$ and
$\epsilon_{k}=k^{2}$.  The sum over $k$ can be performed by means
of the representation of the Bessel function
\begin{equation}
K_{i\nu}(z)=\int_{0}^{\infty}dt\exp(-z\cosh t)\cos(\nu t)
\end{equation}
We have
\begin{equation}
\begin{array}{l}\sum_{k=1}^{\infty}\exp\Big(-kz\cosh(t)-kz^{\prime}\cosh(t^{\prime})\Big)
\cos(k(x_{2}-x_{2}^{\prime}))=
\cr\Re\Big(\exp\Big(-z\cosh(t)-z^{\prime}\cosh(t^{\prime})+i(x_{2}-x_{2}^{\prime})\Big)\cr
\Big(1-\exp(-z\cosh(t)-z^{\prime}\cosh(t^{\prime})+i(x_{2}-x_{2}^{\prime}))\Big)^{-1}\Big)
\end{array}
\end{equation}
Inserting in eq.(60) ($z=by$) and approximating the denominator in
eq.(62) by 1 we obtain an asymptotic estimate for large $y$ and
$y^{\prime}$ . Then,
\begin{equation}\begin{array}{l}
{\cal G}_{N}(x_{0},x_{1},{\bf
x};x_{0}^{\prime},x_{1}^{\prime},{\bf x}^{\prime})- G
(x_{0},x_{1};x_{0}^{\prime},x_{1}^{\prime})\cr
\simeq\frac{8}{\pi^{2}}\cos(x_{2}-x_{2}^{\prime})\int_{0}^{\infty}
dp_{1}\sinh(\pi p_{1})\exp(-p_{1}\vert x_{0}-x_{0}^{\prime}\vert)
K_{ip_{1}}(b\exp(x_{1}))K_{ip_{1}}(b\exp(x_{1}^{\prime}))\end{array}\end{equation}
We can  estimate  the integral over $p_{1}$ if $\vert
x_{0}-x_{0}^{\prime}\vert>\pi$. In such a case inserting the
asymptotic expansion of the Bessel function we obtain
\begin{equation}\begin{array}{l}
{\cal G}_{N}(x_{0},x_{1},{\bf
x};x_{0}^{\prime},x_{1}^{\prime},{\bf x}^{\prime})-
G(x_{0},x_{1};x_{0}^{\prime},x_{1}^{\prime})\cr
\simeq\frac{8}{\pi^{2}}\cos(x_{2}-x_{2}^{\prime})(\vert
x_{0}-x_{0}^{\prime}\vert-\pi)^{-1}
\exp\Big(-b(\exp(x_{1})+\exp(x_{1}^{\prime}))\Big)\end{array}\end{equation}
It follows that ${\cal G}_{N}$  tends exponentially fast to the
Green function for the two-dimensional quantum field theory.
${\cal G}_{N}$ does not vanish at the horizon. We could impose
Dirichlet boundary conditions on ${\cal G}_{N}$ at the horizon.
For example we could  demand that ${\cal G}_{N}$ is vanishing on
the line $x_{1}= u$ (as suggested in \cite{susskind}). We can
construct such a Green function by means of the method of images.
When $u\rightarrow -\infty$ and $x_{1}\rightarrow -\infty$ (still
$x_{1}>u$) then the Dirichlet Green function ${\cal G}_{N}^{D}$
tends to the two-dimensional Dirichlet Green function of the
two-dimensional Laplace operator vanishing on the line $x_{1}=u$.
However, as pointed in \cite{susskind} it is not obvious that we
should impose such a boundary condition at the horizon.

We can derive an integral representation for the Green function
for the n-dimensional torus $S^{1}\times S^{1}\times....\times
S^{1}$ ( then $\epsilon_{k}=k_{1}^{2}+...+k_{n}^{2}$) applying the
formula
\begin{equation}\begin{array}{l}
\exp(-by\cosh(t)\sqrt{k_{1}^{2}+...+k_{n}^{2}})\cr
=\pi^{-\frac{1}{2}}\int_{0}^{\infty} dr
r^{-2}\exp(-\frac{1}{4r^{2}})\exp\Big(-r^{2}y^{2}b^{2}\cosh^{2}(t)(k_{1}^{2}+...
+k_{n}^{2}) \Big)\end{array}\end{equation} Now the sum over
$k_{1},....,k_{n}$ can be performed. It is expressed by a product
of elliptic $\theta$-functions. A precise analysis of such a
formula is still difficult. We shall rely on an approximation
applicable to general compact ${\cal M}$ (for the torus the Weyl
approximation of the spectrum, applied in eq.(67) below is exact).

We estimate  the rhs of eq.(60) for large $x_{1}$ and
$x_{1}^{\prime}$ by means of a simplified argument applicable
when ${\bf x}={\bf x}^{\prime}$ , $x_{1}=x_{1}^{\prime}$, $\vert
u({\bf x})\vert\leq C$. Then,
\begin{equation}\begin{array}{l}
\vert{\cal G}_{N}(x_{0},x_{1},{\bf x};x_{0}^{\prime},x_{1},{\bf
x})- G(x_{0},x_{1};x_{0}^{\prime},x_{1}) \vert\cr \leq
C^{2}\frac{4}{\pi^{2}}\int_{0}^{\infty}dp_{1}\sinh(\pi
p_{1})\exp(-p_{1}\vert x_{0}-x_{0}^{\prime}\vert) \sum_{k\neq
0}\vert K_{ip_{1}}(b\sqrt{\epsilon_{k}}\exp(x_{1}))\vert^{2}
\end{array}\end{equation}
For the behaviour of the rhs of eq.(66) at large $x_{1}$ only
large $k$ in the sum (66) are relevant (the finite sum over $k$ is
decaying exponentially). For large eigenvalues ($\epsilon_{k}\geq
n$ with $n$ sufficiently large) we can apply the Weyl
approximation for the eigenvalues
 distribution \cite{taylor} with the conclusion
\begin{equation}\begin{array}{l}
\vert{\cal G}_{N}(x_{0},x_{1},{\bf x};x_{0}^{\prime},x_{1},{\bf
x})- G(x_{0},x_{1};x_{0}^{\prime},x_{1}) \vert \cr \leq
C^{2}\frac{4}{\pi^{2}}\int_{0}^{\infty}dp_{1}\sinh(\pi
p_{1})\exp(-p_{1}\vert x_{0}-x_{0}^{\prime}\vert) \sum_{\delta
\leq \epsilon_{k}\leq n}\vert
K_{ip_{1}}(b\sqrt{\epsilon_{k}}\exp(x_{1}))\vert^{2}\cr+
R\int_{0}^{\infty}dp_{1}\sinh(\pi p_{1})\exp(-p_{1}\vert
x_{0}-x_{0}^{\prime}\vert) \int_{\vert {\bf k}\vert\geq
\sqrt{n}}d{\bf k}\vert K_{ip_{1}}(b\vert{\bf
k}\vert\exp(x_{1}))\vert^{2}
\end{array}\end{equation}
where $\delta$ is the lowest non-zero eigenvalue.

 The finite sum
as well as the integral on the rhs of eq.(67) are decaying
exponentially. This follows from estimates on Bessel functions
$K_{\nu}(z)$ for large values of $z$ \cite{grad}. The analogous
formula for ${\cal M}=R^{n}$ (see eq.(71) at the end of this
section and ref. \cite{haba}) does not lead to the logarithmic
term $G$ on the lhs of eq.(60) because $\delta=0$ in this case
(there is no gap in the spectrum and no zero mode).

 We introduce now a free Euclidean field as a
random field with the correlation function equal to the Green
function ${\cal G}_{N}$ (see \cite{jaffe}; we need $\int f{\cal
G}_{N}f\geq 0$)\begin{equation}\begin{array}{l}\Phi
(x_{0},x_{1},{\bf x}) =\int dp_{0}dp_{1}\sum_{k}
a_{k}(p_{0},p_{1})\phi_{k}^{p_{1}}(x_{1}) u_{k}({\bf
x})\exp(ip_{0}x_{0})\cr
=\Phi_{0}(x_{0},x_{1})+\sum_{k>0}\Phi_{k}(x_{0},x_{1},{\bf x})
\end{array} \end{equation}where
\begin{equation}
\langle
\overline{a}_{k}(p_{0},p_{1})a_{k^{\prime}}(p_{0}^{\prime},p_{1}^{\prime})\rangle=
\delta_{kk^{\prime}}\delta(p_{0}-p_{0}^{\prime})\delta(p_{1}-p_{1}^{\prime})
(p_{0}^{2}+p_{1}^{2})^{-1}\end{equation}
 $\Phi_{0}$ is two-dimensional conformal invariant free field
with the correlation function
\begin{equation}
\langle\Phi_{0}(x_{0},x_{1})\Phi_{0}(x_{0}^{\prime},x_{1}^{\prime})\rangle
=G(x_{0},x_{1};x_{0}^{\prime},x_{1}^{\prime})=-\frac{1}{4\pi}\ln
\Big((x_{0}-x_{0}^{\prime })^{2}+(x_{1}-x_{1}^{\prime })^{2}\Big)
\end{equation}
  The
correlation functions of $\Phi_{k}$ are decaying exponentially.
This is so, because for large $x_{1}$ the eigenfunctions
$\phi_{k}^{p_{1}}$ are solutions behind the potential barrier.
Hence, the field $\Phi (x_{0},x_{1},{\bf x})$ will decrease
exponentially fast for  $x_{1}\rightarrow +\infty$.

Finally, let us still write down the formulas for ${\cal
M}=R^{D-1}$ (the conventional Rindler space discussed in
\cite{fulling}\cite{boulware}\cite{sciama} and our earlier paper
\cite{haba}). Then, the spectrum of the Laplace-Beltrami operator
on ${\cal M}$ is continuous and instead of the sum over $k$ we
have an integral over the momenta ${\bf p}$
\begin{equation}\begin{array}{l} {\cal G}_{N}(x_{0},x_{1},{\bf
x};x_{0}^{\prime},x_{1}^{\prime},{\bf x}^{\prime})\cr
=(2\pi)^{-D+1}\int dp_{1}d{\bf p}\vert p_{1} \vert^{-1}\exp(-\vert
p_{1}\vert \vert x_{0}-x_{0}^{\prime}\vert) \overline{\phi}_{{\bf
p}}^{p_{1}}(x_{1}^{\prime})\phi_{{\bf p}}^{p_{1}}(x_{1})
\exp(i{\bf p}({\bf x}-{\bf x}^{\prime}))\end{array}\end{equation}
 $\phi_{{\bf p}}^{p_{1}}(x_{1})$ is defined in eq.(57) where $\epsilon_{k}$ is replaced by ${\bf p}^{2}$.
 Then, in order to obtain the formula for the Green function (71) $u_{k}({\bf x})$ in eq.(52)
 are replaced by the plane waves
$(2\pi)^{-\frac{D}{2}+\frac{1}{2}} \exp(i{\bf px})$.

The expansion of the Euclidean  field  takes the form
\begin{equation}\Phi
(x_{0},x_{1},{\bf x}) =(2\pi)^{-\frac{D}{2}+\frac{1}{2}}\int
dp_{1}d{\bf p}a(p_{0},p_{1},{\bf p})\phi_{{\bf p}}^{p_{1}}(x_{1})
\exp(i{\bf px}+ip_{0}x_{0})
\end{equation}and
\begin{displaymath}
\langle \overline{a}(p_{0},p_{1},{\bf
p})a(p_{0}^{\prime},p_{1}^{\prime},{\bf
p}^{\prime})\rangle=\delta({\bf p}-{\bf
p}^{\prime})\delta(p_{0}-p_{0}^{\prime})\delta(p_{1}-
p_{1}^{\prime})(p_{0}^{2}+p_{1}^{2})^{-1}\end{displaymath} We can
continue analytically the Green functions to the real time. The
Green functions define quantum fields if they satisfy some
positivity conditions (reflection positivity on the Riemannian
manifold \cite{dimock}\cite{jaffe}and Wightman positivity on the
pseudoRiemannian manifold). The Euclidean and quantum fields
satisfy the identity (true for Riemannian as well as
pseudoRiemanian metrics)
\begin{equation}
\Phi(x)=-\int
dx^{\prime}\sqrt{g}\Big(\triangle_{N}(x^{\prime}){\cal
G}_{N}(x,x^{\prime})\Big)\Phi(x^{\prime})
\end{equation}
Integrating by parts in eq.(73) we express $\Phi(x)$ by the value
of $\Phi(x)$ on the boundary (at infinity) where according to
eqs.(60) and (68) only the two-dimensional free field $\Phi_{0}$
on $R^{2}$ contributes (because $\Phi_{k}$ vanish at the boundary
for $k>0$) and the field $\triangle_{N}\Phi$ (which is zero if
$\Phi$ is the free massless field continued analytically to the
pseudoRiemannian manifold) . The expression of $\Phi$ by its zero
mode $\Phi_{0}$ indicates that we could construct the quantum
field in the Fock space of the massless two-dimensional quantum
free field. We can obtain a representation of the Virasoro algebra
in this Hilbert space.

The formal analytic continuation of free Euclidean fields
$x_{0}\rightarrow ix_{0}$ (a generalization of the quantum fields
on the Rindler space of refs. \cite{fulling}\cite{boulware}\cite{sciama}) reads
\begin{equation}\begin{array}{l}\Phi
(x_{0},x_{1},{\bf x}) =\int dp_{1}\sum_{k}
a_{k}(p_{1})\phi_{k}^{p_{1}}(x_{1}) u_{k}({\bf x})\exp(-i\vert
p_{1}\vert x_{0})\cr +\int dp_{1}\sum_{k}
a_{k}^{+}(p_{1})\phi_{k}^{p_{1}}(x_{1}) \overline{u_{k}}({\bf
x})\exp(i\vert p_{1}\vert x_{0})
\end{array} \end{equation}
where $a$ and $a^{+}$ are annihilation and creation operators
satisfying the commutation relations
\begin{equation}
[a_{k}(p_{1}),a_{l}^{+}(p_{1}^{\prime})]=\delta_{kl}\delta(p_{1}-p_{1}^{\prime})
\end{equation}

\section{Conclusions} We have shown in sec.5 that if the manifold ${\cal M}$
defining the bifurcate Killing horizon is compact then near the
horizon of ${\cal N}$ the massless Green function of ${\cal N}$
can be approximated by the Green function of the two-dimensional
massless free field.  If the manifold ${\cal M}$ is not compact,
but rather resembles the flat space in its volume growth (see
eq.(A.1)), then we cannot expect the reduction to a
two-dimensional model. We cannot apply the eigenfunction expansion
of sec.5 but we refer to the results of the Appendix. Let us
consider the behaviour at the horizon when
$y^{\prime}=y=\exp(x_{1})\rightarrow 0$, i.e., $x_{1}\rightarrow
-\infty$. In this limit, if simultaneously $\sigma_{M}({\bf
x},{\bf x}^{\prime})<<yy^{\prime}$ (without this assumption the
limiting behaviour of Green functions would not depend on ${\bf
x}$) then for $D=3$ ($k=0$ in eq.(A.22)) for any
$x_{0}-x_{0}^{\prime}$
\begin{equation}
{\cal G}=y^{-1}y^{\prime -1}\sigma(\sinh
\sigma)^{-1}\Big(\sigma^{2}+(x_{0}-x_{0}^{\prime})^{2}\Big)^{-1}
\rightarrow \sigma_{M}^{-2}
\end{equation}
We can show that eqs.(A.22)-(A.23) imply a generalization of
eq.(76) to $D+1$ dimensional manifold ${\cal N}$, where for
$x_{1},x_{1}^{\prime}\rightarrow -\infty$
\begin{equation}{\cal G}_{N}(x_{0}-x_{0}^{\prime},\sigma)
\simeq \sigma_{M}^{-D+1}
\end{equation}
It is interesting that this limiting behaviour does not depend on
$x_{0}$ .

Let us consider the case ${\bf x}\simeq {\bf x}^{\prime}$ in more
detail . Then, $\sigma\simeq \vert x_{1}-x_{1}^{\prime}\vert$ and
from eqs.(A.22)-(A.23) for small $x_{1}$ and $x_{1}^{\prime}$
\begin{equation}
{\cal G}^{0}_{N}(x_{0}-x_{0}^{\prime},\sigma)\simeq
\left(\left(x_{1}-x_{1}^{\prime}\right)^{2}+\left(x_{0}-x_{0}^{\prime}\right)^{2}\right)^{-\frac{D}{2}+\frac{1}{2}}
\end{equation}
For arbitrary $x_{1}$ and $x_{1}^{\prime}$ we have directly from
eq.(A.22)
\begin{equation}\begin{array}{l}{\cal
G}^{0}_{N}(x_{0}-x_{0}^{\prime},\sigma)_{2k+2}
=\exp\left(-\left(k+1\right)\left(x_{1}^{\prime}+x_{1}\right)\right)\Big(\sinh(x_{1}-x_{1}^{\prime}
)^{-1}\frac{d}{dx_{1}}\Big)^{k}\cr(x_{1}-x_{1}^{\prime})
(\sinh(x_{1}-x_{1}^{\prime}))^{-1}
\Big((x_{1}-x_{1}^{\prime})^{2}+(x_{0}-x_{0}^{\prime})^{2}\Big)^{-1}
\end{array}\end{equation} It is decaying exponentially for
$x_{1}\rightarrow +\infty$ and $x_{1}^{\prime}\rightarrow
+\infty$. This behaviour results from the tunnelling through the
barrier described by the quantum mechanical equation (56). We can
obtain from eq.(A.23) the exponential decay in $x_{1}$ also for
the odd case ($n=2k+1$)  but a derivation needs some detailed
estimates.

 The behaviour (77) is different than the one
of eq.(67) (which is suggested in \cite{padma}). We can conclude
that if the Laplace-Beltrami operator on the manifold ${\cal M}$
has a discrete spectrum then the Green function on the manifold
${\cal N}$ close to the bifurcate Killing horizon is approximately
logarithmic like in the two-dimensional massless free field
theory. If however the spectrum of the Laplace-Beltrami operator
on ${\cal M}$ is continuous then such an approximation is invalid.
In this case we obtain a power like behaviour at the horizon (
eq.(77)) as for the conventional D+1 massless Euclidean field
theory in a flat space. Away from the horizon we have an
exponential decay of the Green function ${\cal G}_{N}$
characteristic to the tunneling through the barrier. In the case
of the discrete spectrum of $\triangle_{M}$ the zero mode of the
compact manifold ${\cal M}$ is not damped by the barrier. As a
result its contribution to the Green function ${\cal G}_{N}$ is
separated from the tunneling modes leading to the two-dimensional
behaviour of the Green function at the bifurcate Killing horizon.
The two-dimensional behaviour indicates the relevance of the
infinite dimensional conformal group for manifolds with a
bifurcate Killing horizon (as briefly discussed at the end of
sec.5; it has been shown   first in another way in \cite{carlip}).
A realization of the conformal group in the Fock space of quantum
fields on the horizon is discussed in \cite{moretti}. We intend to
continue a study of the conformal invariance in the way suggested
at the end of sec.5 in
 a forthcoming publication.

\section{Appendix:The Green function of ${\cal D}$ with a power-like
increasing volume of ${\cal M}$}
\renewcommand\theequation{A.\arabic{equation}}
\setcounter{equation}{0} It is difficult to determine the volume
growth $\Omega(r)$ (40) in general. We restrict ourselves to a
class of  manifolds \cite{ricci}( with non-negative Ricci
curvature ) such that
\begin{equation}
aA(n-1)r^{n-1}\leq \omega(r)\leq A(n-1) r^{n-1}
\end{equation}with $0<a\leq 1$.  We show in this Appendix that
the volume growth determines  lower and upper bounds on the Green
functions. The index $n$ in eq.(A.1) does not need to be related
to the dimension of the manifold. We show that the index
determines the behaviour of the heat kernel and the Green
functions ( for  other derivations of some estimates on Green
functions from the volume growth see \cite{vol}). Applying
the inequality (A.1) we can conclude from eq.(44)  (as $P_{\tau}$ and $f$
in eq.(44) are positive)  that
\begin{equation}\begin{array}{l} aA(n-1)\int d\rho
P_{\tau}(\rho+v)(\sqrt{2yy^{\prime}\rho})^{n-2} \leq
\int_{0}^{\infty}P_{\tau}(\rho+v)f(2yy^{\prime}\rho)d\rho\cr
=T_{n}(\tau,v)\leq A(n-1)\int_{0}^{\infty} d\rho P_{\tau}(\rho+v)
(\sqrt{2yy^{\prime}\rho})^{n-2}
 d\rho
\end{array}\end{equation}
We solve the inequalities (A.2) with respect to the heat kernel.
 We denote the solution of the inequality by $p_{\tau}$.
 Then,
 \begin{equation}
 ap_{\tau}\leq P_{\tau}\leq p_{\tau}\end{equation}and for Green functions
\begin{equation}
 a{\cal G}_{H}^{m}\leq {\cal G}_{D}^{m}\leq {\cal G}^{m}_{H}\end{equation}
 When $a=1$ then ${\cal M}=R^{n}$ and the manifold ${\cal D}$ is the hyperbolic space .
 In such a case the heat kernel $p_{\tau}$ and the Green function
 ${\cal G}_{H}$
 are the ones of the  hyperbolic space. Hence, in
 eqs.(A.3)-(A.4) we estimate the heat kernels and Green functions
 on ${\cal D}$(as functions of the invariant distance) by the heat kernels and
 Green functions for the hyperbolic space.

  Let $p_{\tau}$ be the solution of eq.(44) with $f$ defined in eq.(45) resulting from the upper bound (A.2)(the one
  for the hyperbolic space) .
  Then,
  for an even $n=2k$ the equation for $p_{\tau}$ reads
\begin{equation}\begin{array}{l}
2^{k-1}A(n-1)\int_{0}^{\infty} p_{\tau}(\rho+v)\rho^{k-1}d\rho=
\cr (2\pi
 \tau)^{-\frac{1}{2}}\exp\Big(-\frac{1}{2\tau}w^{2}-\frac{n^{2}\tau}{8}\Big)
\equiv F_{n}(v)\end{array}\end{equation} We assume that
$p_{\tau}(x+v)$ is a $k$-th derivative of a certain function
$F_{n}$. Then, integrating by parts in eq.(A.5) we find that this
function is $F_{n}$. More precisely
\begin{equation}
p_{\tau}(v)=(-2\pi)^{-k}\frac{d^{k}}{dv^{k}}F_{2k}(\tau,v)
\end{equation}

The formula for an odd $n=2k+1$ is not so simple. The equation for
$p_{\tau}$ reads
\begin{equation}
2^{k-\frac{1}{2}}A(2k)\int d\rho
p_{\tau}(\rho+v)\rho^{k-\frac{1}{2}}=F_{2k+1}(v)\end{equation}
 First, let us consider $k=0$. We set $\rho=\frac{1}{2}\alpha^{2}$.Then, eq.(A.7) takes the form
\begin{displaymath}
\int_{0}^{\infty} d\alpha
p_{\tau}(\frac{1}{2}\alpha^{2}+v)=F_{1}(\tau,v)
\end{displaymath}
Differentiating over $v$
\begin{displaymath}
\int d\alpha
p_{\tau}^{\prime}(\frac{1}{2}\alpha^{2}+v)=F_{1}^{\prime}(\tau,v)
\end{displaymath}
Shifting $v $ by $\frac{1}{2}\gamma^{2}$ and integrating over
$\gamma$ we obtain
\begin{equation}
\int d\alpha d\gamma
p_{\tau}^{\prime}(\frac{1}{2}(\alpha^{2}+\gamma^{2})+v)=\int
d\gamma F_{1}^{\prime}(\tau,v+\frac{1}{2}\gamma^{2})
\end{equation}
In order to perform the integral on the lhs we introduce  the
cylindrical coordinates $\alpha=r\cos\phi$ and $\gamma=r\sin\phi$.
Then, the integral (A.8) is equal to
\begin{equation}
p_{\tau}(v)=-\frac{1}{\pi}\int_{-\infty}^{\infty}d\gamma
F_{1}^{\prime}(\tau,v+\frac{1}{2}\gamma^{2})
\end{equation}
For $n=2k+1$ we have (here $p_{\tau}^{(0)}=p_{\tau}$)
\begin{equation}
A(2k)\int_{0}^{\infty}p_{\tau}^{(k)}(\frac{1}{2}\alpha^{2}+v)\alpha^{2k}d\alpha=F_{2k+1}(\tau,v)
\end{equation}
Assume that
\begin{equation}
p_{\tau}^{(k)}(v)=\frac{d^{k}}{dv^{k}}h_{k}(\tau,v)
\end{equation}
Then, integrating by parts  we obtain from eq.(A.10)
\begin{equation}
A(2k)(-1)^{k+1}(2k-1)\times(2k-3)\times ...\times
3\int_{0}^{\infty}d\alpha
h_{k}(\tau,\frac{1}{2}\alpha^{2}+v)=F_{2k+1}(\tau,v)
\end{equation}
Hence, we reduced the problem to the one for $k=0$. Eqs.(A.9) and
(A.12) lead to the result
\begin{equation}
p_{\tau}^{(k)}(v)=(-2\pi)^{-k+1}\int_{-\infty}^{\infty}d\gamma\frac{d^{k+1}}{dv^{k+1}}
F_{2k+1}(\tau,\frac{1}{2}\gamma^{2}+v)
\end{equation}
We can  express $p_{\tau}^{(0)}$ ( eq.(A.9)) in the McKean form
\cite{mckean} changing the integration variable $\gamma$ into $r$
where $v+\frac{\gamma^{2}}{2}=\cosh r -1$. Then,
\begin{equation} p_{\tau}^{(0)}(\sigma)=\exp(-\frac{\tau}{8})\sqrt{2}(2\pi
\tau)^{-\frac{3}{2}} \int_{\sigma}^{\infty}(\cosh r -\cosh
\sigma)^{-\frac{1}{2}}r\exp(-\frac{r^{2}}{2\tau})dr
\end{equation}
The solutions (A.6) and (A.13) are equal to the ones for the heat
kernel on the  hyperbolic space. According to eq.(A.4) as
functions of the invariant distance (33) on ${\cal D}$ these
expressions approximate the heat kernel on ${\cal D}$. Let us
write down these formulas in a more explicit form. We have for odd
dimensions $D=n+1=2k+3$ ($k=0,1,..$)
\begin{equation}
p_{\tau}^{(k+1)}(\sigma)=(-2\pi)^{-k}\exp(-\frac{n^{2}}{8}\tau+\frac{1}{2}\tau)
\Big((\sinh\sigma)^{-1}\frac{d}{d\sigma}\Big)^{k}p_{\tau}^{(1)}(\sigma)
\end{equation}
with \begin{equation} p_{\tau}^{(1)}(\sigma)=(2\pi
\tau)^{-\frac{3}{2}}\sigma(\sinh\sigma)^{-1}\exp(-\frac{\tau}{2}-\frac{\sigma^{2}}{2\tau})
\end{equation}
In even dimensions $D=n+1=2k+2$
\begin{equation}
\begin{array}{l}
p_{\tau}^{(k)}(\sigma)=2\exp(-\frac{n^{2}\tau}{8}+\frac{\tau}{8})(-2\pi)^{-k}
\Big((\sinh \sigma)^{-1}\frac{d}{d\sigma}\Big)^{k}
p_{\tau}^{(0)}(\sigma)
\end{array} \end{equation}These formulas have been derived in another way in
 \cite{rycher}\cite{anker}\cite{grig}\cite{campo}.

 Then, the massless Green function on the hyperbolic space is ($k=0,1,...$)\begin{equation}
{\cal G}_{H}^{m}(y,{\bf x};y^{\prime},{\bf
x}^{\prime})_{2k+2}=(-2\pi)^{-k} \Big((\sinh
\sigma)^{-1}\frac{d}{d\sigma}\Big)^{k}\sigma(\sinh
\sigma)^{-1}\exp(-\nu\sigma)
\end{equation}
where \begin{equation}
\nu=\sqrt{\frac{n^{2}}{4}+m^{2}}\end{equation}
 We obtain analogous formulas for the Green
functions  for an odd $n$
  \begin{equation} \begin{array}{l}{\cal
G}_{D}^{m}(y,{\bf x};y^{\prime},{\bf x}^{\prime})_{2k+1}\cr
 =2\sqrt{2}(2\pi)^{-\frac{3}{2}}
(-2\pi)^{-k} \Big((\sinh \sigma)^{-1}\frac{d}{d\sigma}\Big)^{k}
\int_{\sigma}^{\infty}(\cosh r -\cosh
\sigma)^{-\frac{1}{2}}\exp(-\nu r)dr\cr =2(2\pi)^{-\frac{3}{2}}
(-2\pi)^{-k} \Big((\sinh
\sigma)^{-1}\frac{d}{d\sigma}\Big)^{k}Q_{\nu-\frac{1}{2}}(\cosh
\sigma)
\end{array}
\end{equation}
where (\cite{grad})
\begin{displaymath}
 Q_{\mu}(\cosh \sigma)=\int_{\sigma}^{\infty}(2\cosh r -2\cosh
\sigma)^{-\frac{1}{2}}\exp(-\frac{(2\mu+1)r}{2})dr
\end{displaymath}
The Fourier transform in $x_{0}$ of the massless  Green function
(6) on ${\cal N}$ is equal to the  Green function on ${\cal D}$
with (see eq.(21))
\begin{displaymath}
\nu=\vert p_{0}\vert
\end{displaymath}
in eq.(A.18),i.e., for an odd dimension
\begin{equation}
\tilde{{\cal G}}_{N}^{0}(p_{0},y,{\bf x};y^{\prime},{\bf
x}^{\prime})_{2k+2}=(-2\pi)^{-k}y^{-k-1}y^{\prime-k-1} \Big((\sinh
\sigma)^{-1}\frac{d}{d\sigma}\Big)^{k}\sigma(\sinh
\sigma)^{-1}\exp(-\vert p_{0}\vert\sigma)
\end{equation}The massless  Green function in D+1 dimensional manifold ${\cal
N}$ , when $D=n+1=2(k+1)+1$, can be obtained either from eq.(A.21)
by means of the Fourier transform or from eq.(27) by a calculation
of the $\tau$-integral
\begin{equation}
{\cal G}^{0}_{N}(x_{0}-x_{0}^{\prime},\sigma)_{2k+2}
=y^{-k-1}y^{\prime -k-1}\Big((\sinh
\sigma)^{-1}\frac{d}{d\sigma}\Big)^{k}\sigma
(\sinh\sigma)^{-1}\Big(\sigma^{2}+(x_{0}-x_{0}^{\prime})^{2}\Big)^{-1}
\end{equation}

 In even
dimensions $D=2k+2$ the formula is more complicated . From
eq.(A.20) we obtain
\begin{equation}\begin{array}{l} {\cal G}^{0}_{N}(x_{0}-x_{0}^{\prime},\sigma)_{2k+1} =
y^{-\frac{k+1}{2}}y^{\prime -\frac{k+1}{2}}\Big((\sinh
\sigma)^{-1}\frac{d}{d\sigma}\Big)^{k}\cr
\int_{\sigma}^{\infty}(\cosh r -\cosh
\sigma)^{-\frac{1}{2}}r(r^{2}+(x_{0}-x_{0}^{\prime})^{2})^{-1}dr
\end{array}\end{equation}
In order to obtain approximate formulas for the generalized
hyperbolic manifold ${\cal D}$ and the Rindler-type manifold
${\cal N}$ with the horizon we should insert in eqs.(A.14)-(A.23)
the expression for the invariant distance $\sigma$ (33) on ${\cal
D}$.

\end{document}